\documentclass[mathleft
]{an}
\usepackage{graphicx}
\usepackage{times}
\newcommand{\fsec}{\hbox{$.\!\!{\arcsec}$}}
\newcommand{\kms}{{\,km\,s$^{-1}$}}

\begin{document}

% The following seven commands are intended for editorial usage and should be
% ignored by the author(s).
\Pagespan{1}{}% Document's page range. 
% If second parameter is left empty, the last page is computed automatically.
\Yearpublication{}%
\Yearsubmission{}%
\Month{}%   
\Volume{}%  
\Issue{}% 
% \DOI{This.is/not.aDOI}% 

\title{Photospheric and chromospheric activity on EY~Dra\,\thanks{Based on 
    observations made with the Nordic Optical Telescope, operated on the 
    island of La Palma jointly by Denmark, Finland, Iceland, Norway, and 
    Sweden, in the Spanish Observatorio del Roque de los Muchachos of the 
    Instituto de Astrofisica de Canarias.}}

\author{H. Korhonen\inst{1}\fnmsep\thanks{Corresponding author: 
    \email{hkorhonen@aip.de}}%\newline} 
  \and 
  K. Brogaard \inst{2}
  \and        
  K. Holhjem \inst{3,4} 
  \and   
  S. Ramstedt \inst{5} 
  \and  
  J. Rantala \inst{6}
  \and  
  C.C. Th{\"o}ne \inst{7} 
  \and 
  K. Vida \inst{8,9}} 
\titlerunning{Photospheric and chromospheric activity on EY~Dra}
\authorrunning{Korhonen et al.}
\institute{
Astrophysikalisches Institut Potsdam, An der Sternwarte 16, 
D-14482 Potsdam, Germany
    \and
     Department of Physics and Astronomy, University of Aarhus,
     DK-8000 Aarhus C, Denmark
     \and
     Nordic Optical Telescope, Apartado 474, E-38700 Santa Cruz de La Palma,
     Spain
     \and 
     Argelander-Institut f{\"u}r Astronomie, Auf dem H{\"u}gel 71, D-53121
     Bonn, Germany 
     \and
     Stockholm Observatory, AlbaNova University Centre, S-10691 Stockholm,
     Sweden
     \and
     Observatory, PO Box 14, FI-00014 University of Helsinki, Finland
     \and
     Dark Cosmology Centre, Niels Bohr Institute, University of Copenhagen,
     Juliane Maries Vej 30, DK-2100 Copenhagen {\O}, Denmark
     \and
     E{\"o}tv{\"o}s Lor{\'a}nd University, Department of Astronomy, H-1518
     Budapest, PO Box 32, Hungary
     \and
     Konkoly Observatory of the Hungarian Academy of Sciences, H-1525
     Budapest, Hungary 
   }

\received{}
\accepted{}
\publonline{}


\abstract{Magnetic activity in the photosphere and chromosphere of the M
  dwarf EY~Dra is studied and possible correlations between the two are
  investigated using photometric observations in the V and R bands and
  optical and near infrared spectroscopy. The longitudinal spot configuration
  in the photosphere is obtained from the V band photometry, and the
  chromospheric structures are investigated using variations in the H$\alpha$
  line profile and observations of the Paschen $\beta$ line. The shape of the
  V band light-curve indicates two active regions on the stellar surface, about
  0.4 in phase apart. The spectroscopic observations show enhanced H$\alpha$ 
  emission observed close to the phases of the photometrically detected
  starspots. This could indicate chromospheric plages associated with the
  photospheric starspots. Some indications of prominence structures are also
  seen. The chromospheric pressure is limited to $\log {\rm m}_{\rm TR} < -4$
  based on the non-detection of emission in the Paschen~$\beta$ wavelength
  region.}

\maketitle

\section{Introduction}

EY~Dra (RE~1816+541) is a very active M dwarf that was discovered by the 
ROSAT extreme ultraviolet (EUV) all sky survey in the early 1990's (Pounds
et~al.~\cite{pounds}). The optical counterpart of the EUV source RE~1816+541
was first observed by Jeffries et~al.~(\cite{jef_etal}) who did a thorough
analysis of the object using optical spectroscopy. They observed strong,
variable H$\alpha$ emission and molecular lines, and concluded that the source
is an M1-2e dwarf, and thereby one of the most active stars in the solar
neighbourhood. They also discovered that EY~Dra is a very rapid rotator with  
a $v\sin i\approx 61$\kms\ and a rotation period of about 12~$\sin i$
hours. Jeffries et~al.~(\cite{jef_etal}) determined the radial velocity of the
target to be $-21.9\pm 1.5$\kms. A higher spectral and temporal resolution
study of EY~Dra was carried out by Eibe (\cite{eibe}) who found significant
variations in the H$\alpha$ profile, which were interpreted as chromospheric
plages and prominence clouds higher up in the atmosphere, but still below the
corotation radius. 

The first photometric observations of EY~Dra were obtained by Schwartz 
et~al.~(\cite{schw_etal}) who established the star's brightness to be 11.83 in
the V band and the colours of (B-V)=1.45, (V-R)=0.96 and (R-I)=1.05, also 
indicating a cool star. The first long-term photometric study of EY~Dra was 
carried out by Robb~\& Cardinal~(\cite{robb_card}) who measured the rotation
period of $0\fd 4589$ and remarked that the light-curve shape indicated two
large spots or active regions on the stellar surface. Barnes~\& Collier
Cameron~(\cite{barn_col}) used Doppler imaging techniques to obtain the
first precise surface structure maps of EY~Dra. These surface temperature maps
showed spots on a very large latitude range ($20^{\circ} - 80^{\circ}$), but
no polar spot. 

In this paper we carry out the first simultaneous photometric and spectroscopic
observations of EY~Dra to study the photospheric spots and correlate them with
the variability seen in the chromosphere. The longitudinal spot configuration
is obtained from the photometric observations and the chromosphere is studied
with the high resolution H$\alpha$ line observations. In addition, medium
resolution near infrared (NIR) observations of the Paschen $\beta$ line were
used to further investigate the chromosphere of EY~Dra.

\section{Observations}

All the observations presented in this paper were obtained at the Nordic
Optical Telescope in early July 2006. The optical data were taken using
ALFOSC, which is owned by the Instituto de Astrofisica de Andalucia (IAA) and
operated at the Nordic Optical Telescope under agreement between IAA and the
NBIfAFG of the Astronomical Observatory of Copenhagen. The NIR spectroscopy
was obtained using NOTCam.  

All the observations have been phased using the ephem\-eris, ${\rm HJD} =
2449927.752 + 0\fd 4589 \times {\rm E}$, from Robb \& Cardinal
(\cite{robb_card}). 

\subsection{Photometry}

Photometric observations of EY~Dra in the V and R bands were obtained with
ALFOSC. The detector is an E2V Technologies 2k back-illuminated CCD with
13.5$\mu$ pixels, giving a field-of-view (FOV) of $6.5^{\prime} \times
6.5^{\prime}$. This FOV made it possible to observe the comparison star
(GSC 0390400259) and the check star (GSC 0390400361) in one frame together
with the target, and thus do differential photometry of EY Dra. The comparison
star and the check star are both from the Hubble Space Telescope Guide Star
Catalogue (Jenkner et~al.~\cite{jenk_etal}). 

The photometric observations of EY~Dra were obtained during the nights
starting 2006 July 1 and July 3. The exposure time was 1--3 seconds in the V
band and 1 second in the R band. In total observations from 9 rotational
phases were obtained. Each observation consisted of 1-7 individual exposures
that were first bias and sky flat field corrected, and then averaged after the
determination of the differential magnitude (object-comparison star). The
error of each phase was determined as the standard deviation of all the points
used in the average and divided by the square-root of the number of
observations. The data reduction was done and the photometry obtained using
Image Reduction and Analysis Facility (IRAF) distributed by
KPNO/NOAO. Table~\ref{ph_data} gives more details on the photometric
observations.  

\begin{table}
\caption{Details of the photometric observations of EY~Dra. The Heliocentric 
  Julian Date, rotational phase, instrumental differential magnitude,
  the error of the magnitude and the number of observations used to obtain 
  the magnitude are given for both the V and R bands. The error for
  each data point is the standard deviation of the measurements
  divided by the square root of the number of the measurements.} 
\label{ph_data}
\begin{center}     
\begin{tabular}{c c c l c}
\hline
HJD & phase & mag & error & no \\
2453000+ &  &   &         &      \\
\hline 
\multicolumn{5}{c}{V band}\\
918.41440 & 0.148 & -1.197 & 0.002 & 4 \\
918.46784 & 0.265 & -1.202 & 0.010$^{*}$ & 1 \\
918.52188 & 0.382 & -1.184 & 0.004 & 3 \\
918.61168 & 0.578 & -1.163 & 0.003 & 3 \\
920.41944 & 0.517 & -1.173 & 0.004 & 5 \\
920.53116 & 0.761 & -1.172 & 0.003 & 2 \\
920.58829 & 0.885 & -1.163 & 0.004 & 3 \\
920.62317 & 0.961 & -1.149 & 0.007 & 2 \\
920.67216 & 0.068 & -1.178 & 0.003 & 3 \\
\multicolumn{5}{c}{R band} \\
918.41845 & 0.157 & -1.772 & 0.003 & 4 \\
918.46948 & 0.268 & -1.773 & 0.003 & 3 \\ 
918.52399 & 0.387 & -1.757 & 0.004 & 3 \\
918.61382 & 0.583 & -1.738 & 0.003 & 3 \\
920.42230 & 0.524 & -1.749 & 0.004 & 3 \\
920.54907 & 0.800 & -1.718 & 0.010$^{*}$ & 1 \\
920.59153 & 0.892 & -1.732 & 0.002 & 7 \\
920.62466 & 0.965 & -1.729 & 0.004 & 3 \\
920.67405 & 0.072 & -1.749 & 0.002 & 3 \\
\hline
\end{tabular}
\end{center}
*) Only one observation was obtained during this phase, so no standard
 deviation of the measurements could be obtained to estimate the error of the
 data point. An error value of 0.010 was adopted.
\end{table}

\subsection{Optical spectroscopy}

Optical spectroscopy of EY~Dra around the H$\alpha$ line was obtained using
ALFOSC, grism\#17, and a $0\fsec 5$ off-set slit during the nights starting
2006 July 1 and July 3. This instrument configuration gives a resolving power
($\lambda / \Delta\lambda$) of 10~000 and a spectral coverage approximately
from 6200~{\AA} to 6700~{\AA}. Due to the fringing in the E2V CCD starting
around 6400~{\AA}, the observations were done in sets of five 120 second
exposures. Between each separate spectrum the object was moved along the
slit to be able to remove this fringe pattern from the observations. After
every five object exposures, two Halogen flat fields and one Neon arc
spectrum were obtained. After basic reduction steps (bias subtraction, image
trimming and flat field correction) the five consecutive observations were
combined to obtain 15 better signal-to-noise (S/N) spectra with minimum fringe
patterns. A radial velocity standard (HD~103095) and a B star (BD+33 2642)
were also observed. The B star spectrum was used for checking contribution
from terrestrial lines in this spectral region. The reductions were carried
out using the 4A reduction package (Ilyin~\cite{ilyin}). More details on the
observations are given in Table~\ref{sp_data}. 

\begin{table}
\caption{Optical spectroscopy of EY~Dra. Heliocentric Julian Date, rotational 
  phase, radial velocity obtained with IRAF fxcor task and the S/N per pixel
  are given.} 
\label{sp_data}
\centering     
\begin{tabular}{c c c c }
\hline
HJD & phase & RV & S/N \\
2453000+ &  & [\kms ]   &     \\
\hline 
918.44256 & 0.210 & $-21.6 \pm 2.7$ & 150 \\
918.45631 & 0.239 & $-21.3 \pm 3.0$ & 162 \\
918.48694 & 0.306 & $-17.7 \pm 3.6$ & 128 \\
918.49864 & 0.332 & $-17.6 \pm 3.3$ & 94 \\
918.50970 & 0.356 & $-15.6 \pm 3.1$ & 93 \\
918.62720 & 0.612 & $-28.7 \pm 3.3$ & 58 \\
920.39421 & 0.462 & $-25.3 \pm 2.5$ & 116 \\
920.56335 & 0.831 & $-22.7 \pm 2.1$ & 139 \\
920.57345 & 0.853 & $-15.9 \pm 2.3$ & 164 \\
920.64873 & 0.017 & $-13.4 \pm 2.3$ & 133 \\
920.66079 & 0.043 & $-13.5 \pm 2.2$ & 114 \\
920.69449 & 0.117 & $-31.8 \pm 2.6$ & 131 \\
920.70800 & 0.146 & $-33.3 \pm 2.3$ & 108 \\
920.71966 & 0.172 & $-31.9 \pm 2.5$ & 124 \\
920.72899 & 0.192 & $-33.7 \pm 2.0$ & 91 \\
\hline   
\end{tabular}
\end{table}

\subsection{Near IR spectroscopy}

The medium resolution NIR observations were obtained in the region around the
Paschen~$\beta$ line, using the NOTCam with the high resolution camera,
grism\#1, and J filter. This instrument configuration gives a resolving power
($\lambda / \Delta\lambda$) of 5700 and a wavelength coverage of
12620--13520~{\AA}. The detector is a  Rockwell Science Center "HAWAII" array
with $1024 \times 1024 \times 18.5$~$\mu$m pixels in HgCdTe. Two spectra with
$4 \times 450$~sec exposure time were obtained in the evening of 2006 July
5. 

For removing the IR background we used an ABBA dithering pattern along the
slit, which gave 4 separate spectra that were combined into one spectrum
during the reductions. Each spectrum was obtained using non-destructive
readouts which were acted upon by a linear regression calculation reducing the
Poisson noise of the observation. The whole spectral region shows strong
fringing pattern that did not completely disappear even after the flat
fielding. For removing the skylines an A0 star (HD~172728) was observed as an
atmospheric standard. The data were reduced using IRAF. 

\section{Radial velocity}

The radial velocity of EY~Dra was investigated from the lines in the
wavelength region 6400--6500~{\AA} using the IRAF fxcor routine. Before the 
cross-correlation all three radial velocity standard observations were
combined to one higher S/N spectrum, after which the resulting spectrum was
spun-up to the $v\sin i$ of EY~Dra, 61~\kms\ (Jeffries
et~al.~\cite{jef_etal}).  The measurements for the individual phases are given
in Table~\ref{sp_data}. 

The radial velocity for the whole dataset is $-22.6 \pm 1.9$~\kms . This is in
agreement with the values published by (Jeffries et~al.~\cite{jef_etal}) and
Eibe (\cite{eibe}). The radial velocities obtained from each night's data set
are $-20.1 \pm 1.9$~\kms\  and $-24.3 \pm 2.9$~\kms , for 2006 July 1 and July
3, respectively. Note that the errors stated assume that the errors are
random.

\section{Photosphere}

\begin{figure}
  \centering
  \includegraphics[width=0.40\textwidth]{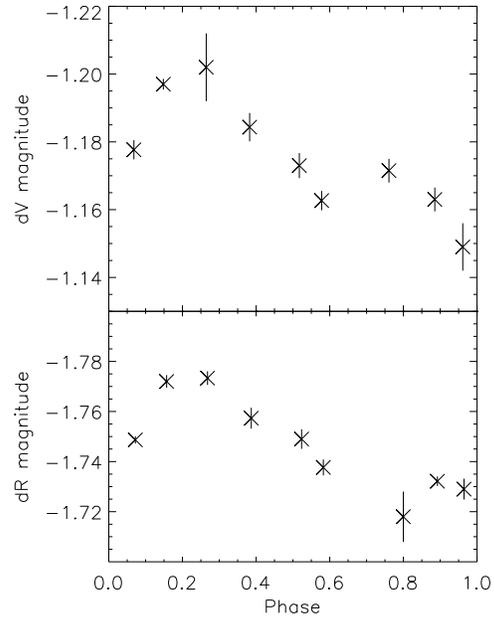}
  \caption{Differential V and R band photometry of EY~Dra with the errors.} 
  \label{EYDra_ph}
\end{figure}

The V and R band light-curves of EY~Dra together with the errors of the
individual points are shown in Fig.~\ref{EYDra_ph}. The V band light-curve has
in general a W shape. The phases 0.76 and 0.88 form the bump seen in the broad
light-curve minimum. The two minima in the V band light-curve are located
around the phases 0.3--0.75 and 0.85--1.15. The main maximum occurs around
the phases 0.15--0.3. On the whole, the V band light-curve indicates two
active regions separated by about 0.4 in phase on the surface of EY~Dra.

\begin{figure}
  \centering
  \includegraphics[width=0.45\textwidth]{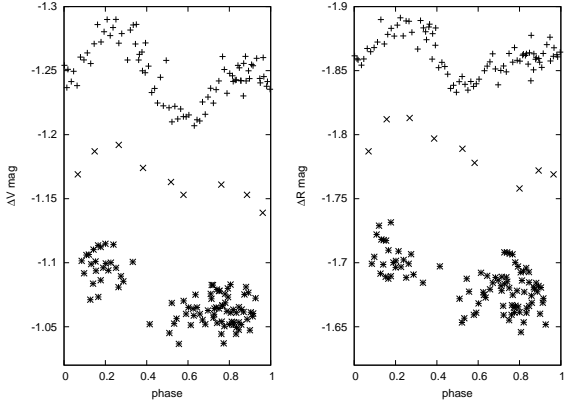}
  \caption{Differential V (left) and R (right) band observations of 
  EY~Dra. The plus-signs are data obtained at the Konkoly observatory on
  average 62 days before the NOT observations, crosses are the NOT 
  observations and the stars are observations obtained at the Konkoly 
  observatory on average 27 days after the NOT observations. Offsets have been
  applied to the individual datasets to show the light-curve shapes better.}
  \label{EYDra_kon}
\end{figure}

In the V band observations of Robb~\& Cardinal~(\cite{robb_card}) from 1995
similar W shape was present. Also, the photometry obtained before and after
our observations at the Konkoly observatory (Vida~\& Ol{\'a}h
\cite{vida_olah}) exhibits the W shape. This photometry, which will be
published later, is shown in Fig.~\ref{EYDra_kon} together with our
observations. In the plot the plus-signs are data obtained at the Konkoly
observatory between 2006 April 21 and May 12, crosses are the NOT observations
and the stars are observations from the Konkoly observatory obtained between
2006 July 22 and August 8. The relatively large scatter seen in the latter
Konkoly observations is most likely due to the non-optimal observing
conditions during this time period.

In the R band observations obtained at NOT the W shape is not really seen. The
light-curve shows a broad minimum around the phases 0.8--1.1. The observations
at the phase 0.89 show indications of the bump, but unfortunately the phase
0.8, which also would be in the bump, only has one individual observation and
as such a large photometric error. The R band observations from the Konkoly
observatory show the W shape for the time periods before and after the NOT
observations. When the instrumental (V-R) colour is calculated, no clear
modulation of the colour with the stellar rotation period is seen. But the
data point at the phase 0.8 deviates strongly from the others. 

\section{Chromosphere}

\subsection{H$\alpha$ line}

\begin{figure}
  \centering
  \includegraphics[width=0.40\textwidth]{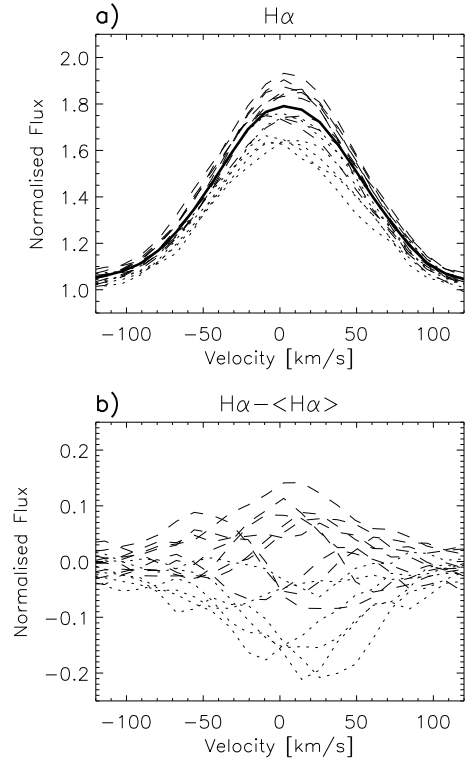}
  \caption{The H$\alpha$ line variations in EY~Dra. a) All the 15 individual
  H$\alpha$ observations obtained in this study. The line profiles are plotted
  against the velocity obtained in relation to the rest wavelength of the
  H$\alpha$. The radial velocity has been removed from all the profiles. The
  thick line gives the average line profile. b) The residual variations in the
  H$\alpha$. The mean profile has been subtracted from all the 
  observations. In both of the plots the dotted line gives the observations
  from the night starting 2006 July 1 and the dashed line observations 
  from the night starting 2006 July 3.}
  \label{EYDra_Ha}
\end{figure}

In EY~Dra the H$\alpha$ line is a quite broad emission feature. All the
profiles from the 15 epochs observed in this study are plotted in
Fig.~\ref{EYDra_Ha}{\bf a}. The radial velocities stated in
Table~\ref{sp_data} have been removed from the profiles. As can be seen, the
strength of the profile varies significantly with time. The thick line in the
plot is the mean of all the profiles. The residual variations in the profile
after the mean profile has been subtracted are given in
Fig.~\ref{EYDra_Ha}{\bf b}. For both plots in Fig.~\ref{EYDra_Ha} the data
from the night starting 2006 July 1 are presented with dotted lines and the
data from the night starting 2006 July 3 are given by dashed lines. 

\begin{figure}
  \centering
  \includegraphics[width=0.45\textwidth]{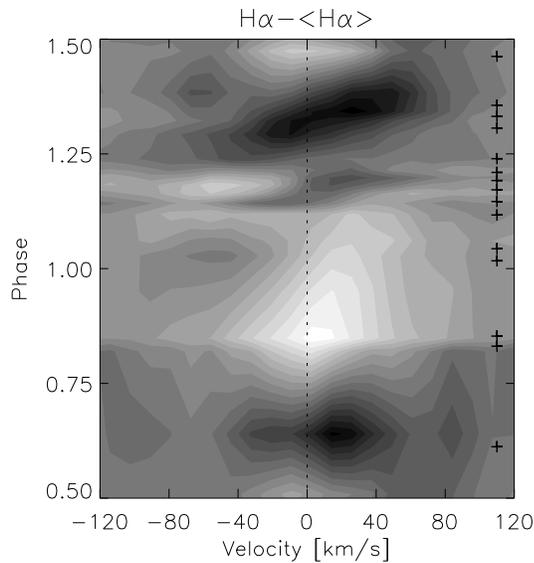}
  \caption{Dynamic spectrum of the H$\alpha$ line of EY~Dra. The image shows 
  the difference profiles after the mean profile has been subtracted from the 
  observations. Bright colour means more emission in the H$\alpha$. The 
  measured radial velocity for each observations has been removed from the
  profiles. The crosses on the right hand side of the plot give the phases of
  the observations and the dashed line the 0 velocity.} 
  \label{EYDra_dyn}
\end{figure}

To study the H$\alpha$ behaviour in more detail the profiles showing the
difference between the observations and the mean spectrum were investigated
more thoroughly. In Fig.~\ref{EYDra_dyn} a dynamic spectrum constructed from
these difference profiles is shown. Brighter colours in the plot correspond
to enhanced emission and the darker colours to the emission that is less than
the average. The phases of the observations are shown with crosses on the
plot. The data for the phases where there are no observations are just
interpolations between the closest phases with data. 

The relatively low spectral resolution of the data does not allow such a
detailed analysis of the chromospheric absorption and emission clouds as was
done in Eibe~(\cite{eibe}). However, our data also indicates chromospheric
structures. The dynamic spectrum clearly shows enhanced emission in the
H$\alpha$ around the phases 0.75--1.1, and a small enhancement around the
phase 1.5. In the dynamic spectrum the features with increased and decreased
H$\alpha$ emission often seem to move from blue to red, and could indicate
chromospheric plages and prominences. Especially the enhanced emission around
the phases 0.75--1.1 seems to move from blue to red, and could be associated
with a plage. The other enhancement, around the phase 0.5, is, due to the
sparse data sampling, only seen in one spectrum. The feature could be
associated with a plage, though this cannot be confirmed. The feature of
decreased H$\alpha$ emission around the phases 1.25--1.5 is seen to move
across the line profile, and could thus be associated with a prominence
cloud. It has to be noted though, that the interpolation used to fill the
phase gaps in the dynamic spectrum can artificially enhance the perception of
moving features. 

\subsection{Paschen $\beta$ line}

Atmospheric models of M dwarfs calculated by Short~\&
Doyle~(\cite{short_doyle}) show that the Paschen $\beta$ line profile can be
used to determine the chromospheric pressure and thereby the activity level in
such stars. This spectral line, which is in the NIR part of the spectrum
($\lambda$=12~818 {\AA}), was therefore also observed.

\begin{figure}
  \centering
  \includegraphics[width=0.40\textwidth]{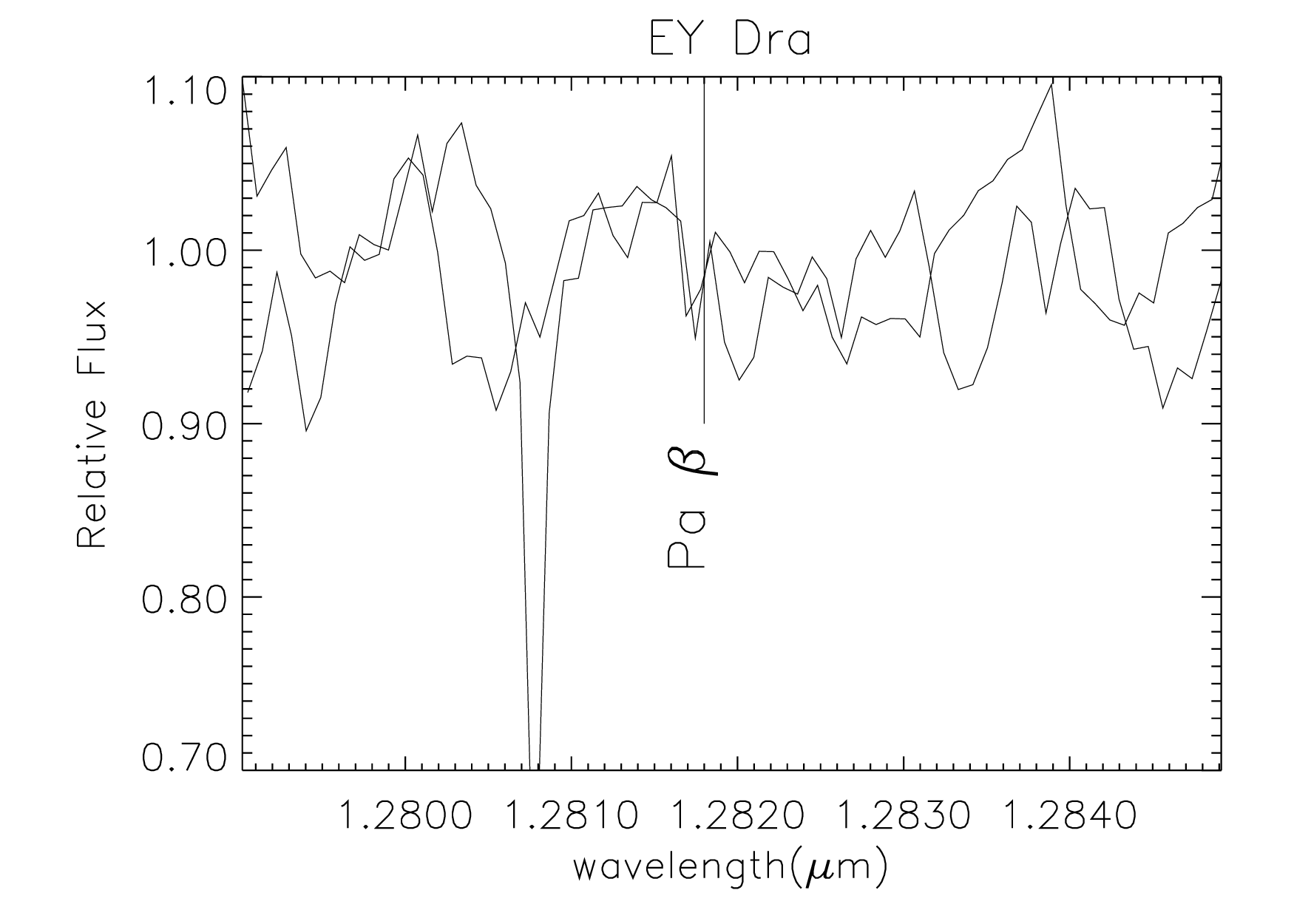}
  \caption{The final two background subtracted, flat fielded, normalised and 
    radial velocity corrected NIR spectra of EY Dra. The absorption line 
    present in one spectrum and not in the other is a telluric feature. The 
    vertical line is the position where the Paschen $\beta$ line centre should 
    be if present in the spectra. } 
  \label{EYDra_IR}
\end{figure}

Fig.~\ref{EYDra_IR} shows the two observed NIR spectra of EY~Dra around the
Paschen $\beta$ line. The resolution of the spectra is poor and a strong
fringing pattern is seen in the spectra even after the flat field correction.
Still, it is clear from the observations that no strong emission line is
seen around the Paschen $\beta$ wavelength. 

\section{Discussion}

\subsection{Night to night variation of the H$\alpha$} 

\begin{figure}
  \centering
  \includegraphics[width=0.45\textwidth]{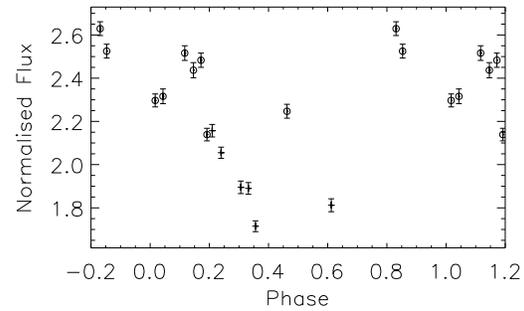}
  \caption{The variation of the H$\alpha$ flux obtained by integrating over
  the whole line and plotted against the rotational phase. The observations
  from the night starting 2006 July 1 are given by crosses and the
  observations from the night starting July 3 by circles.} 
  \label{EYDra_Haph}
\end{figure}

When looking at Fig.~\ref{EYDra_Ha} it seems that in general the observations
obtained during the night starting 2006 July 1 (dotted line) show less
emission than the ones observed during the night starting 2006 July 3 (dashed
line). This could be caused by a real difference in the activity level between
the two nights, or by the observations from the first night coinciding with
the less active rotational phases. In Fig.~\ref{EYDra_Haph} the integrated
H$\alpha$ flux is plotted against the phase. The behaviour seen during the
second observing night (circles), can be interpreted as more or less constant  
H$\alpha$ emission, whereas the behaviour during the first observing night
(crosses), shows clearly diminishing activity with increasing phase. 

In principal the behaviour seen during the first observing night could be
explained by a gradual decline of a chromospheric flare. On the other hand, if
the behaviour seen during the first observing night is due to a declining
flare, then flux seen in the phases 0.3--0.35 indicates the normal H$\alpha$
flux. This would imply that the observations from the second observing night,
that have a higher flux than that, would be from a flaring state. Also, if it
indeed is a flare that is seen during the second observing night, then this
flare would last the whole night without showing any decline. We also note
that the H$\alpha$ flux at the phase 0.192, from the second observing night,
is almost identical with the flux at the phase 0.210, obtained during the
first observing night (see Fig.~\ref{EYDra_Haph}). All this implies that the
difference seen in the activity levels between these two nights is most likely
just caused by the observations during the first observing night coinciding
with the less active regions on the stellar surface.

\subsection{Spot longitudes}
\label{inv}

For measuring the longitudes of the photospheric spots a spot filling factor
map was obtained from the V band light-curve using light-curve inversion
methods (see e.g., Lanza et al.~(\cite{lanzaetal}) for the method and Ol{\'a}h
et al.~(\cite{olahetal}) for the implementation used here). Light-curve
inversions were decided to be used for the spot longitude determination,
as tests show that the inversions result in more accurate determination of the
spot longitudes than just simply determining the light-curve minimum gives (see
Savanov \& Strassmeier \cite{sav_str}). In some cases, especially when
dealing with close-by spots, taking the light-curve minimum results in wrong
spot longitude, whereas the inversion gives the correct longitudinal spot
configuration. 

For the inversion the unspotted surface temperature is set to 4000~K, which is
consistent with the observations of the spectral type. Spots are assumed to be
1000~K cooler than the unspotted surface, in line with observations of other
active stars. The instrumental differential V band magnitude that
corresponds to the unspotted surface is estimated to be $-1\fm 25$, though as
the photometric time series of EY~Dra is short this parameter is relatively
uncertain. However, changing the brightest magnitude will not affect the
positions of the spots seen in the filling factor maps, only the filling
factor values themselves. The inclination was set to 66\degr\ measured by
Robb~\& Cardinal~(\cite{robb_card}).

One should also note that one-dimensional data, as the light-curve is, do not
give information on the latitudinal distribution of the spots. This means that
the spot latitude seen in the maps arises from the fact that the inversion
process tends to introduce the spots to the location where they have the
maximum impact on the light-curve, i.e., at the centre of the visible stellar
disk. From photometry it is impossible to discern whether the light-curve
minimum is caused by a single large spot or an active region consisting of
several spots. For simplicity the structure causing the light-curve minimum,
and seen in the spot filling factor map, is called a spot. 

\begin{figure}
  \centering
  \includegraphics[width=0.40\textwidth]{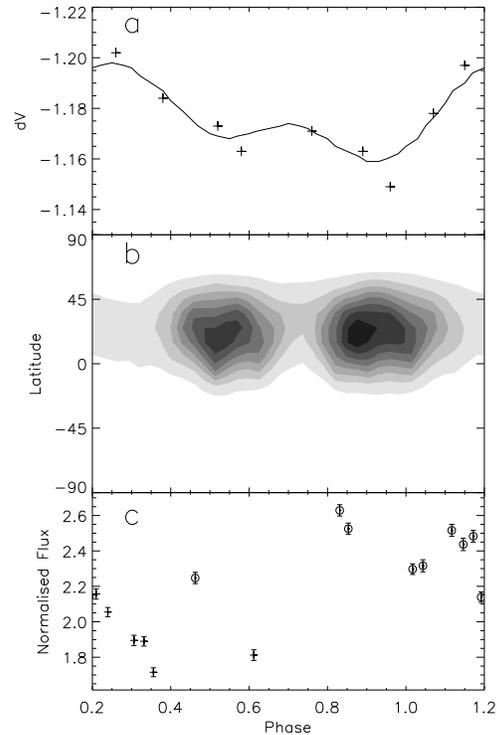}
  \caption{Correlating photospheric and chromospheric activity. a)  
  The differential V band observations (crosses) with the fit (solid line)
  obtained from the inversion. b) The spot filling factor map, where the
  darker colour means larger spot filling factor. c) Variation in the
  H$\alpha$ flux (symbols as in Fig.~\ref{EYDra_Haph}). In all the plots the
  phases from 0.2 to 1.2 are plotted to show the active regions better.}
  \label{EYDra_inv}
\end{figure}

The resulting filling factor map of EY~Dra for early July 2006 is given in
Fig.~\ref{EYDra_inv}{\bf b}. It clearly shows two large spots on the
surface. These spots are located at phases 0.4--0.6 (centred at the
phase 0.53) and 0.8--1.1 (centred at the phase 0.91). This implies two
active longitudes separated by 0.4 in phase. A longer time series of
photometric observations is needed for confirming whether or not this
configuration is the normal case for EY~Dra. The earlier photometric
observations (Robb~\& Cardinall~\cite{robb_card}; Vida~\&
Ol{\'a}h~\cite{vida_olah}) have also shown the W shape, which implies
that this kind of active longitude structure is relatively stable on
EY~Dra. A spot configuration where two spots are located on the
stellar surface about 0.5 in phase apart is common for active stars
(see e.g., Jetsu et al.~\cite{jetsuetal}; Berdyugina~\&
Tuominen~\cite{ber_tuo}). Recent dynamo calculations can also produce 
active longitudes that are 0.25--0.5 in phase apart (e.g., Moss \cite{moss};
Elstner~\& Korhonen~\cite{elst_kor}). 

\subsection{Correlating activity in the photosphere and chromosphere}

In the Sun, photospheric dark spots are often associated with bright plages in
the chromosphere. Some active stars also show evidence for correlated magnetic
activity in the photosphere and chromosphere. Alekseev \& Kozlova
(\cite{ak_lqhya}) investigated quasi-simultaneous photometric observations,
photo-polarimetry and high-resolution spectra of a young solar-like star
LQ~Hya. They found evidence for a connection between plages, magnetic regions
and the starspot longitudes. Similar results have been obtained also for an
active binary MS~Ser (Alekseev \& Kozlova \cite{ak_msser}). 

The possible correlation between the photospheric spots and chromospheric
plages on EY~Dra was studied using the data presented in this paper. The spot
positions determined from the light-curve inversions of the V band data
(Fig.~\ref{EYDra_inv}{\bf a,b}) show that the photospheric active regions
occur at two active longitudes: one centred around the phase 0.53 and
the other one centred around the phase 0.91. On the other hand the
enhanced H$\alpha$ emission occurs at the phases 0.8--1.1 and another
small increase is seen around the phase 0.5 (see
Fig.~\ref{EYDra_inv}{\bf c}). Unfortunately the extent of this weaker 
feature cannot be determined, as there is only one observation around 
this enhancement. In general, the increases in the H$\alpha$ emission
occur near the same phases where the photospheric spots are centred. Thus, the
observations of EY~Dra obtained at NOT in early July 2006 can be interpreted
as bright plages in the chromosphere associated with the photospheric active
region centred at the phase 0.91, and possibly also on the photospheric active
region centred at the phase 0.53.

\subsection{Chromospheric pressure}

The Paschen $\beta$ line is expected to vary with chromospheric pressure;
changing with increasing pressure from a weak absorption line to a stronger
absorption line and finally to an emission line (Short~\&
Doyle~\cite{short_doyle}). The pressure at the top of the chromosphere, or 
equivalently at the bottom of the transition region, is measured as a column
density and given by $\log {\rm m}_{TR}$. According to the models by Short~\&
Doyle~(\cite{short_doyle}) the line responds to the increasing chromospheric
pressure in the same way as the H$\alpha$. For low chromospheric pressure it
is weakly in absorption with minimal equivalent width $W_\lambda$. As the
chromospheric pressure increases the line becomes more strongly absorbent, 
with maximal $W_\lambda$ occurring at $\log {\rm m}_{TR} = -4.2$. Then, as the
chromospheric pressure increases further the line makes a rapid transition to
emission. With $\log {\rm m}_{TR} = -4.0$ the line is either balanced
between absorption and emission or is weakly in emission, depending on the
exact model. The more precise behaviour can be seen in Sort~\&
Doyle~(\cite{short_doyle}, Fig.~3) 

Our NIR spectra of EY~Dra do not show any indication of an emission line
around the wavelength of Paschen~$\beta$. According to the models of Short~\& 
Doyle~(\cite{short_doyle}) the chromospheric pressure of EY~Dra is hence $\log
{\rm m}_{TR} \leq -4$. The non-detection of the Paschen~$\beta$ line in the
EY~Dra spectra can be due to the low resolution of the observations, though a
strong emission line would be visible even with this relatively low quality
data. 

Short~\& Doyle~(\cite{short_doyle}) also present models for the H$\alpha$
line. According to their calculations the pressure value of $\log {\rm m}_{TR}
= -4.0$ causes the H$\alpha$ to be strongly in emission, even though the
Paschen $\beta$ line is not yet in emission. The H$\alpha$ emission observed
in EY~Dra supports the idea that the chromospheric pressure of EY~Dra is close
to $\log {\rm m}_{TR} = -4.0$, as with much lower pressures the H$\alpha$ line
would not be in emission either. 

\section{Conclusions}

Based on the photometric and spectroscopic observations of EY~Dra the
following conclusions can be drawn:

\begin{itemize}
\item The light-curve shape indicates two active regions approximately 0.4 in
  phase apart. Other photometric observations of EY~Dra show that this
  could be the normal spot configuration. 
\item The H$\alpha$ shows strong variations during the two nights of
  observations. Indications for plages and prominences are seen.
\item The main enhancement seen in the H$\alpha$ emission occurs close to the
  phases of the photospheric active region that is centred at the phase
  0.91. This indicates bright chromospheric plages associated with the dark
  photospheric spots, as is often seen in the Sun. 
\item Chromospheric pressure is limited to $\log {\rm m}_{\rm TR} < -4$, based
  on the non-detection of Paschen~$\beta$ emission.
\end{itemize}

\begin{acknowledgements}
  The observations used in this paper were obtained during the 2006 Nordic
  Baltic Research School at NOT and SST in La Palma, running from June 27
  until July 8 2006. This summer school was financed by NordForsk and the main 
  organiser was Dr.\ J.\ Sollerman. The authors would also like to thank 
  Dr.\ A.\ Djupvik for her extremely valuable help with the NIR observations
  and reductions and Dr.\ M.I.\ Andersen for his help with the photometry. HK 
  acknowledges the support from the German \emph{Deut\-sche
  For\-schungs\-ge\-mein\-schaft, DFG\/} project number KO 2310/1-2. KH
  acknowledges support from a doctoral fellowship awarded by the Research
  council of Norway, project number 177254/V30. KV acknowledges the financial
  support of the Hungarian government through OTKA T048961 and T043504. The
  Dark Cosmology Centre is funded by the Danish National Research Foundation.
\end{acknowledgements}

\end{document}